# A SIMPLE DERIVATION OF THE EINSTEIN-MAXWELL FIELD EQUATIONS FROM THE $2^{nd}$ ORDINARY EXTERIOR DIFFERENTIAL OF A PRECURSOR TO THE SOLDERING FORM

C. C. Briggs
*Center for Academic Computing, Penn State University, University Park, PA 16802*

**Abstract.** A simple derivation is given of the Einstein-Maxwell field equations from the $2^{nd}$ ordinary exterior differential of a precursor to the soldering form for $n$-dimensional differentiable manifolds having a general linear connection and in 5-dimensional general relativity in particular.

This paper presents a simple derivation of the Einstein-Maxwell field equations from the $2^{nd}$ ordinary exterior differential of a precursor to the soldering form for $n$-dimensional differentiable manifolds $M$ having a general linear connection and in 5-dimensional general relativity in particular.

The products

$$\mathrm{e}_a \otimes \omega^b = \mathrm{e}_a\, \omega^b = \mathrm{e}_c\, \delta^c_a\, \delta^b_d\, \omega^d \tag{1}$$

of the basis tangent vectors $\mathrm{e}_a$ and basis 1-forms $\omega^a$ of $M$ are basis elements[1] of the tensor product[2] $T(M) \otimes T^*(M)$ of the tangent and cotangent bundles $T(M)$ and $T^*(M)$ of $M$ and are a precursor to "the soldering form"[3-5] (also called "the Cartan-Maurer form,"[6] "the canonical form,"[7-8] and "the displacement vector"[9-10]) $\mathrm{d\,P}$ given by

$$\mathrm{d\,P} = \mathrm{e}_a \otimes \omega^a = \mathrm{e}_a\, \omega^a = \mathrm{e}_a\, \delta^a_b\, \omega^b, \tag{2}$$

where $\delta^a_b$ is the Kronecker delta and where Latin indices represent anholonomic coordinates.

The $1^{st}$ ordinary exterior differentials of $\mathrm{e}_a$ are given by[11-13]

$$\mathrm{d\,e}_a = \mathrm{e}_b\, \omega_a{}^b, \tag{3}$$

the contractions of which with $\omega^b$ are given by

$$\langle \omega^b, \mathrm{d\,e}_a \rangle = \omega_a{}^b \tag{4}$$

and in view of which the $1^{st}$ absolute exterior differentials of $\mathrm{e}_a$ are given by

$$\begin{aligned} \mathrm{D\,e}_a &= \mathrm{d\,e}_a - \mathrm{e}_b\, \omega_a{}^b \\ &= 0, \end{aligned} \tag{5}$$

where $\omega_a{}^b$ is the connection 1-form of $M$ as given in terms of the connection coefficients $\Gamma_a{}^b{}_c$ by[14-19]

$$\omega_a{}^b = \Gamma_c{}^b{}_a\, \omega^c, \tag{6}$$

and where the contractions of $\omega^b$ with $\mathrm{e}_a$ are given by

$$\langle \omega^b, \mathrm{e}_a \rangle = \delta^b_a. \tag{7}$$

The $2^{nd}$ ordinary exterior differentials of $\mathrm{e}_a$ are given by[20-29]

$$\begin{aligned} \mathrm{d}^2\, \mathrm{e}_a &= \mathrm{d\,d\,e}_a \\ &= \mathrm{d\,e}_b\, \omega_a{}^b \\ &= (\mathrm{d\,e}_b) \wedge \omega_a{}^b + \mathrm{e}_b\, \mathrm{d}\, \omega_a{}^b \\ &= \mathrm{e}_c\, \omega_b{}^c \wedge \omega_a{}^b + \mathrm{e}_b\, \mathrm{d}\, \omega_a{}^b \\ &= \mathrm{e}_b\, (\mathrm{d}\, \omega_a{}^b + \omega_c{}^b \wedge \omega_a{}^c) \\ &= \mathrm{e}_b\, \Omega_a{}^b, \end{aligned} \tag{8}$$

the contractions of which with $\omega^b$ are given by

$$\langle \omega^b, \mathrm{d}^2\, \mathrm{e}_a \rangle = \Omega_a{}^b, \tag{9}$$

where $\Omega_a{}^b$ is the curvature 2-form of $M$ as given by

$$\begin{aligned} \Omega_a{}^b &= \mathrm{d}\, \omega_a{}^b + \omega_c{}^b \wedge \omega_a{}^c \\ &= \mathrm{D}\, \omega_a{}^b + \omega_a{}^c \wedge \omega_c{}^b \\ &= \mathrm{d}\, \omega_a{}^b - \omega_a{}^c \wedge \omega_c{}^b \\ &= \mathrm{D}\, \omega_a{}^b - \omega_c{}^b \wedge \omega_a{}^c \\ &= \mathrm{d}\, \omega_a{}^b + \tfrac{1}{2}\, [\omega_c{}^b, \omega_a{}^c] \\ &= \mathrm{D}\, \omega_a{}^b + \tfrac{1}{2}\, [\omega_a{}^c, \omega_c{}^b] \\ &= \mathrm{d}\, \omega_a{}^b - \tfrac{1}{2}\, [\omega_a{}^c, \omega_c{}^b] \\ &= \mathrm{D}\, \omega_a{}^b - \tfrac{1}{2}\, [\omega_c{}^b, \omega_a{}^c] \\ &= \tfrac{1}{2}\, R_{cda}{}^b\, \omega^c \wedge \omega^d, \end{aligned} \tag{10}$$

---

1 Goldberg, S. I., *Curvature and Homology*, Dover Publications, Inc., New York, NY (1982), p. 9.
2 Chern, S. S., "Geometry of characteristic classes," in Vanstone, J. R., (ed.), *Proceedings of the Thirteenth Biennial Seminar of the Canadian Mathematical Congress on Differential Topology; Differential Geometry and Applications, Vol. 1*, Canadian Mathematical Congress, Montreal, Canada (1972), p. 1.
3 Chern, S. S., "Integral Formulas for the Characteristic Classes of Sphere Bundles," *Proc. Nat. Acad. Sci.*, **30** (1944) 269; "Some Theorems on the Isometric Imbedding of Compact Riemann Manifolds in Euclidean Space," *Ann. of Math.*, **56** (1952) 422.
4 Misner, C. W., K. S. Thorne, and J. A. Wheeler, *Gravitation*, W. H. Freeman & Co., San Francisco, CA (1973), p. 353.
5 Percacci, R., "Spontaneous Soldering," *Phys. Lett.*, **144B** (1984) 37; "Role of Soldering in Gravity Theory," in Doebner, H. J., and T. D. Palev (eds.), *Proceedings of the XIII International Conference on Differential Geometric Methods in Theoretical Physics*, World Scientific Publishing Co. Pte. Ltd., Singapore (1986), p. 250.
6 Chern, S. S., "The Geometry of G-Structures," *Bull. Amer. Math. Soc.*, **72** (1966) 167.
7 Kobayashi, S., and K. Nomizu, *Foundations of Differential Geometry, Vol. 1*, John Wiley & Sons, Inc., New York, NY (1963), p. 140.
8 Choquet-Bruhat, Y., C. DeWitt-Morette, and M. Dillard-Bleick, *Analysis, Manifolds, and Physics*, North-Holland Publishing Co., Amsterdam, The Netherlands (1982), p. 376.
9 Flanders, H., "Development of an Extended Exterior Differential Calculus," *Trans. Amer. Math. Soc.*, **75** (1953) 311.
10 Goldberg, S. I., *op. cit.*, p. 34.
11 Misner, C. W., *et al.*, *op. cit.*, p. 350.
12 Cartan, É. J., *Leçons sur la géométrie des espaces de Riemann*, Gauthier-Villars et Cie, Paris, France (1928), p. 180; *Leçons sur la géométrie des espaces de Riemann, 2nd édition, rev. et augm.*, Gauthier-Villars, Paris, France (1946), p. 180.
13 Chern, S. S., "A simple intrinsic proof of the Gauss-Bonnet formula for closed Riemannian manifolds," *Ann. of Math.*, **45** (1944) 747; "On the curvatura integra in a Riemannian manifold," *ibid.*, **46** (1945) 674; "On the Minimal Immersions of the Two-sphere in a Space of Constant Curvature," in Gunning, R. C. (ed.), *Problems in Analysis: A Symposium in Honor of Salomon Bochner*, Princeton University Press, Princeton, NJ (1970), p. 27.
14 Schouten, J. A., *Ricci-Calculus, Second Edition*, Springer-Verlag, Berlin, Germany (1954), p. 172.
15 Mielke, E. W., "Generating function for new variables in general relativity and Poincaré gauge theory," *Phys. Lett.*, **149A** (1990) 345.
16 Misner, C. W., *ey al.*, *op. cit.*, p. 350.
17 Cartan, É. J., (1928), *op. cit.*, p. 50; (1946), *op. cit.*, p. 48.
18 Nakahara, M., *Geometry, Topology, and Physics*, IOP Publishing Ltd., Bristol, England (1990), p. 245.
19 Kobayashi, S., and K. Nomizu, *op. cit.*, p. 141.
20 Bott, R., and S. S. Chern, "Some Formulas Related to Complex Transgression," in Haefliger, A., and R. Narasimhan (eds.), *Essays on Topology and Related Topics; Memoires dédiés à Georges de Rham*, Springer-Verlag, New York, NY (1970), p. 48.
21 Bradlow, S., "Curvatures," in *The Geometry of Vector Bundles and an Introduction to Gauge Theory*, [unpublished lecture notes], 1998, Lecture 20.
22 Cartan, É. J., "Sur les variétés à connexion affine et la théorie de la relativité généralisée (Première Partie)," *Annales scientifiques de l'École Normale supérieure*, **40** (1923) 325; (1928), *op. cit.*, p. 182; (1946), *op. cit.*, p. 181.
23 Chern, S. S., "Differential Geometry of Fiber Bundles," in *Proceedings of the International Congress of Mathematicians*, **2** (1950) 397; "Topics in Differential Geometry," [mimeographed notes], The Institute for Advanced Study, Princeton, NJ (1951), Chap. 3, Sec. 3; (1970), *op. cit.*
24 Cohen, J. M., *Exterior Differential Forms and Orthonormal Cartan Frames*, [unpublished monograph], (1975), p. 33.
25 Flanders, H., (1953), *op. cit.*; "Methods in Affine Connection Theory," *Pac. J. Math.*, **5** (1955) 391; *Differential Forms with Applications to the Physical Sciences*, Academic Press, New York, NY (1963), pp. 130 and 144.
26 Misner, C. W., *et al.*, *op. cit.*, pp. 350-351 and 359.
27 Petrie, T., and J. Randall, *Connections, Definite Forms, and Four-manifolds*, Clarendon Press, Oxford, England (1990), pp. 11-12.
28 Sattinger, D. H., and O. L. Weaver, *Lie Groups and Algebras with Applications to Physics, Geometry, and Mechanics*, Springer-Verlag, New York, NY (1986), p. 102.
29 Trautmann, A., *Differential Geometry for Physicists—Stony Brook Lectures*, Bibliopolis, edizioni di filosofia e scienze, Napoli, Italy (1984), pp. 91-92.

where $R_{abc}{}^{d}$ is the Riemann-Christoffel curvature tensor of $M$.

Schouten's $2^{nd}$ and $3^{rd}$ identities for $R_{abc}{}^{d}$ are given by[30]

$$R_{[abc]}{}^{d} = 2\,(\nabla_{[a}\,S_{bc]}{}^{d} - 2\,S_{[ab}{}^{e}\,S_{c]e}{}^{d}) = 2\,(\nabla_{[a}\,S_{bc]}{}^{d} + 2\,S_{[ab}{}^{e}\,S_{|e|c]}{}^{d}) \tag{11}$$

and[31]

$$R_{ab(cd)} = \nabla_{[a}\,Q_{b]cd} + S_{ab}{}^{e}\,Q_{ecd}, \tag{12}$$

respectively, where $S_{ab}{}^{c}$ is the torsion tensor and $Q_{abc}$ the non-metricity tensor of $M$, $S_{ab}{}^{c}$ and $Q_{abc}$ being given by

$$S_{ab}{}^{c} = \Gamma_{[a}{}^{c}{}_{b]} + \Omega_{a}{}^{c}{}_{b} \tag{13}$$

and by

$$Q_{a}{}^{bc} = \nabla_{a}\,g^{bc} \tag{14}$$

or equivalently

$$Q_{abc} = -\,\nabla_{a}\,g_{bc} = g_{bd}\,g_{ce}\,Q_{a}{}^{de}, \tag{15}$$

respectively, where $\Omega_{a}{}^{c}{}_{b}$ is the object of anholonomity (which vanishes for a holonomic coordinate system) as given (1) in terms of $A_{a}^{\alpha}$ and $A_{\alpha}^{a}$ by[32]

$$\Omega_{a}{}^{c}{}_{b} = A_{a}^{\alpha}\,A_{b}^{\beta}\,\partial_{[\alpha}\,A_{\beta]}^{c} = -\,A_{\alpha}^{c}\,\partial_{[a}\,A_{b]}^{\alpha} = A_{[b}^{\alpha}\,\partial_{a]}\,A_{\alpha}^{c}, \tag{16}$$

where Greek indices represent holonomic coordinates and where $A_{a}^{\alpha}$ and $A_{\alpha}^{a}$ are defined by the relations

$$\mathrm{e}_{a} = A_{a}^{\alpha}\,\partial_{\alpha} = A_{a}^{\alpha}\,\frac{\partial}{\partial x^{\alpha}} = \partial_{a} \tag{17}$$

and

$$\omega^{a} = A_{\alpha}^{a}\,dx^{\alpha}, \tag{18}$$

respectively, and satisfy the equations

$$A_{a}^{\alpha}\,A_{\beta}^{a} = \delta_{\beta}^{\alpha} \tag{19}$$

and

$$A_{\alpha}^{a}\,A_{b}^{\alpha} = \delta_{b}^{a}, \tag{20}$$

whence

$$A_{a}^{\alpha} = \langle \omega^{\alpha}, \mathrm{e}_{a}\rangle = \langle \mathrm{e}_{a}, \omega^{\alpha}\rangle \tag{21}$$

and

$$A_{\alpha}^{a} = \langle \omega^{a}, \mathrm{e}_{\alpha}\rangle = \langle \mathrm{e}_{\alpha}, \omega^{a}\rangle, \tag{22}$$

and (2) in terms of the commutation coefficents $c_{a}{}^{c}{}_{b}$ by

$$\Omega_{a}{}^{c}{}_{b} = -\tfrac{1}{2}\,c_{a}{}^{c}{}_{b}, \tag{23}$$

while $c_{a}{}^{c}{}_{b}$ can be expressed in terms of the commutator $[\mathrm{e}_{a}, \mathrm{e}_{b}]$ of $\mathrm{e}_{a}$ and $\mathrm{e}_{b}$ defined by

$$\begin{aligned}[\mathrm{e}_{a}, \mathrm{e}_{b}] &= 2\,\partial_{[a}\,\partial_{b]} \\ &= 2\,A_{[a}^{\alpha}\,\partial_{|\alpha|}\,A_{b]}^{\beta}\,\partial_{\beta} \\ &= 2\,A_{[a}^{\alpha}\,(\partial_{|\alpha|}\,A_{b]}^{\beta})\,A_{\beta}^{c}\,\partial_{c} \\ &= -\,2\,A_{[a}^{\alpha}\,A_{b]}^{\beta}\,(\partial_{\alpha}\,A_{\beta}^{c})\,\partial_{c} \\ &= c_{a}{}^{c}{}_{b}\,\mathrm{e}_{c}\end{aligned} \tag{24}$$

as

$$c_{a}{}^{c}{}_{b} = \langle \omega^{c}, [\mathrm{e}_{a}, \mathrm{e}_{b}]\rangle, \tag{25}$$

whence

$$\Omega_{a}{}^{c}{}_{b} = -\tfrac{1}{2}\,c_{a}{}^{c}{}_{b} = -\tfrac{1}{2}\,\langle \omega^{c}, [\mathrm{e}_{a}, \mathrm{e}_{b}]\rangle, \tag{26}$$

while the $1^{st}$ ordinary exterior differentials of $\omega^{a}$ are given by

$$\mathrm{d}\,\omega^{a} = \Omega_{b}{}^{a}{}_{c}\,\omega^{b}\wedge\omega^{c}, \tag{27}$$

whence

$$\Omega_{b}{}^{a}{}_{c} = \tfrac{1}{2!}\,\langle \mathrm{e}_{b}\wedge\mathrm{e}_{c}, \mathrm{d}\,\omega^{a}\rangle, \tag{28}$$

and while the $2^{nd}$ ordinary exterior differentials of $\omega^{a}$ are given by

$$\begin{aligned}\mathrm{d}^{2}\,\omega^{a} &= \mathrm{d}\,\Omega_{b}{}^{a}{}_{c}\,\omega^{b}\wedge\omega^{c} \\ &= (\mathrm{d}\,\Omega_{b}{}^{a}{}_{c})\wedge\omega^{b}\wedge\omega^{c} + \Omega_{b}{}^{a}{}_{c}\,(\mathrm{d}\,\omega^{b})\wedge\omega^{c} - \\ &\quad -\,\Omega_{b}{}^{a}{}_{c}\,\omega^{b}\wedge\mathrm{d}\,\omega^{c} \\ &= (\mathrm{d}\,\Omega_{b}{}^{a}{}_{c})\wedge\omega^{b}\wedge\omega^{c} + \Omega_{b}{}^{a}{}_{c}\,\Omega_{d}{}^{b}{}_{e}\,\omega^{d}\wedge\omega^{e}\wedge\omega^{c} - \\ &\quad -\,\Omega_{b}{}^{a}{}_{c}\,\Omega_{d}{}^{c}{}_{e}\,\omega^{b}\wedge\omega^{d}\wedge\omega^{e} \\ &= (\partial_{[b}\,\Omega_{c}{}^{a}{}_{d]} - 2\,\Omega_{[b}{}^{e}{}_{c}\,\Omega_{d]}{}^{a}{}_{e})\,\omega^{b}\wedge\omega^{c}\wedge\omega^{d} \\ &= 0\end{aligned} \tag{29}$$

using Poincaré's theorem for scalar-valued exterior differential forms,[33-34] i.e.,

$$\mathrm{d}^{2}\,\alpha = 0, \tag{30}$$

where $\alpha$ is an arbitrary scalar-valued exterior differential form, whence[35]

$$\partial_{[b}\,\Omega_{c}{}^{a}{}_{d]} = 2\,\Omega_{[b}{}^{e}{}_{c}\,\Omega_{d]}{}^{a}{}_{e}. \tag{31}$$

In view of Eq. (12) above,

$$R_{abcd} = -\,R_{abdc} + 2\,(\nabla_{[a}\,Q_{b]cd} + S_{ab}{}^{e}\,Q_{ecd}) \tag{32}$$

and

$$\begin{aligned}R_{abc}{}^{c} &= g^{cd}\,R_{abcd} \\ &= g^{cd}\,R_{ab(cd)} \\ &= g^{cd}\,(\nabla_{[a}\,Q_{b]cd} + S_{ab}{}^{e}\,Q_{ecd}) \\ &= \nabla_{[a}\,Q_{b]c}{}^{c} - Q_{[a}{}^{cd}\,Q_{b]cd} + S_{ab}{}^{d}\,Q_{dc}{}^{c} \\ &= \nabla_{[a}\,Q_{b]c}{}^{c} + S_{ab}{}^{c}\,Q_{cd}{}^{d}.\end{aligned} \tag{33}$$

Incidentally, Bianchi's identities for $\Omega_{a}{}^{b}$ are given by[36]

$$\mathrm{D}\,\Omega_{a}{}^{b} = \mathrm{d}\,\Omega_{a}{}^{b} - \omega_{a}{}^{c}\wedge\Omega_{c}{}^{b} + \omega_{c}{}^{b}\wedge\Omega_{a}{}^{c} = 0, \tag{34}$$

the coefficients of which are Bianchi's identities for $R_{abc}{}^{d}$ as given by[37]

$$\tfrac{1}{2}\,\nabla_{[a}\,R_{bc]d}{}^{e} - S_{[ab}{}^{f}\,R_{c]fd}{}^{e} = \tfrac{1}{2}\,\nabla_{[a}\,R_{bc]d}{}^{e} + S_{[ab}{}^{f}\,R_{|f|c]d}{}^{e} = 0 \tag{35}$$

or equivalently by

$$\nabla_{[a}\,R_{bc]d}{}^{e} = 2\,S_{[ab}{}^{f}\,R_{c]fd}{}^{e}. \tag{36}$$

---

[30] Schouten, J. A., *op. cit.*, p. 144.
[31] Schouten, J. A., *op. cit.*, p. 145.
[32] Schouten, J. A., *op. cit.*, p. 100.
[33] Cartan, É. J., (1928), *op. cit.*, pp. 209 and 215; *Le systèmes différentiels extérieurs et leurs applications géometriques*, Actualités scientifiques et industrielles, 994, Exposés de géométrie, vol. 12, Librairie scientifiques Hermann & C$^{ie}$, Éditeurs, Paris, France (1945), p. 37; (1946), *op. cit.*, pp. 208 and 211.
[34] Chern, S. S., "The Geometry of Isotropic Surfaces," *Annals of Math.*, **43** (1942) 545; "On the Euclidean Connections in a Finsler Space," *Proc. Nat. Acad. Sci.*, **29** (1943) 33; "Local Equivalence and Euclidean Connections in Finsler Spaces," *Science Reports Nat. Tsing Hua Univ.*, **5** (1948) 95.
[35] Schouten, J. A., *op. cit.*, pp. 101 and 120.
[36] Cartan, É. J., (1928), *op. cit.*, p. 214; (1946), *op. cit.*, p. 210.
[37] Schouten, J. A., *op. cit.*, p. 146.

The $0^{th}$ ordinary exterior differentials of $\mathrm{e}_a\, \omega^b$ are given by

$$\begin{aligned} \mathrm{d}^0\, \mathrm{e}_a\, \omega^b &= \mathrm{e}_a\, \omega^b \\ &= \mathrm{e}_c\, \delta^c_a\, \delta^b_d\, \omega^d. \end{aligned} \tag{37}$$

The $1^{st}$ ordinary exterior differentials of $\mathrm{e}_a\, \omega^b$ are given by

$$\begin{aligned} \mathrm{d}\, \mathrm{e}_a\, \omega^b &= (\mathrm{d}\, \mathrm{e}_a) \wedge \omega^b + \mathrm{e}_a\, \mathrm{d}\, \omega^b \\ &= \mathrm{e}_c\, \omega_a{}^c \wedge \omega^b + \mathrm{e}_c\, \delta^c_a\, \mathrm{d}\, \omega^b \\ &= \mathrm{e}_c\, (\Gamma_d{}^c{}_a\, \delta^b_e + \delta^c_a\, \Omega_d{}^b{}_e)\, \omega^d \wedge \omega^e \\ &= \mathrm{e}_c\, (\Gamma_{[d}{}^c{}_{|a|}\, \delta^b_{e]} + \delta^c_a\, \Omega_d{}^b{}_e)\, \omega^d \wedge \omega^e. \end{aligned} \tag{38}$$

The $2^{nd}$ ordinary exterior differentials of $\mathrm{e}_a\, \omega^b$ are given by

$$\begin{aligned} \mathrm{d}^2\, \mathrm{e}_a\, \omega^b &= \mathrm{d}\, \mathrm{d}\, \mathrm{e}_a\, \omega^b \\ &= \mathrm{d}\, (\mathrm{e}_c\, \omega_a{}^c \wedge \omega^b + \mathrm{e}_a\, \mathrm{d}\, \omega^b) \\ &= (\mathrm{d}\, \mathrm{e}_c) \wedge \omega_a{}^c \wedge \omega^b + \mathrm{e}_c\, (\mathrm{d}\, \omega_a{}^c) \wedge \omega^b - \mathrm{e}_c\, \omega_a{}^c \wedge \mathrm{d}\, \omega^b + \\ &\quad + (\mathrm{d}\, \mathrm{e}_a) \wedge \mathrm{d}\, \omega^b + \mathrm{e}_a\, \mathrm{d}^2\, \omega^b \\ &= \mathrm{e}_d\, \omega_c{}^d \wedge \omega_a{}^c \wedge \omega^b + \mathrm{e}_c\, (\mathrm{d}\, \omega_a{}^c) \wedge \omega^b - \mathrm{e}_c\, \omega_a{}^c \wedge \mathrm{d}\, \omega^b + \\ &\quad + \mathrm{e}_c\, \omega_a{}^c \wedge \mathrm{d}\, \omega^b + 0 \\ &= \mathrm{e}_c\, (\mathrm{d}\, \omega_a{}^c + \omega_d{}^c \wedge \omega_a{}^d) \wedge \omega^b \\ &= \mathrm{e}_c\, \Omega_a{}^c \wedge \omega^b \end{aligned} \tag{39}$$

or alternatively and more simply by

$$\begin{aligned} \mathrm{d}^2\, \mathrm{e}_a\, \omega^b &= (\mathrm{d}^2\, \mathrm{e}_a) \wedge \omega^b + \mathrm{e}_a\, \mathrm{d}^2\, \omega^b \\ &= \mathrm{e}_c\, \Omega_a{}^c \wedge \omega^b + 0 \\ &= \mathrm{e}_c\, \Omega_a{}^c \wedge \omega^b. \end{aligned} \tag{40}$$

Thus, the $2^{nd}$ ordinary exterior differentials of $\mathrm{e}_a\, \omega^b$ are given by

$$\begin{aligned} \mathrm{d}^2\, \mathrm{e}_a\, \omega^b &= \mathrm{e}_c\, \Omega_a{}^c \wedge \omega^b \\ &= \tfrac{1}{2}\, \mathrm{e}_c\, \delta^b_d\, R_{efa}{}^c\, \omega^d \wedge \omega^e \wedge \omega^f \\ &= \tfrac{1}{2}\, \mathrm{e}_c\, \delta^b_{[d}\, R_{ef]a}{}^c\, \omega^d \wedge \omega^e \wedge \omega^f \\ &= \tfrac{1}{2}\, \mathrm{e}_c\, R^{(3)b}{}_{defa}{}^c\, \omega^d \wedge \omega^e \wedge \omega^f \\ &= \mathrm{e}_c\, G^{(3)b}{}_{defa}{}^c\, \omega^d \wedge \omega^e \wedge \omega^f, \end{aligned} \tag{41}$$

where the coefficients $G^{(3)b}{}_{defa}{}^c$ are defined by

$$\begin{aligned} G^{(3)b}{}_{defa}{}^c &= \tfrac{1}{2}\, R^{(3)b}{}_{defa}{}^c \\ &= \tfrac{1}{2}\, \delta^b_{[d}\, R_{ef]a}{}^c \\ &= \tfrac{1}{3!}\, \langle \mathrm{e}_d \wedge \mathrm{e}_e \wedge \mathrm{e}_f,\, \Omega_a{}^c \wedge \omega^b \rangle \\ &= \tfrac{1}{3!}\, \langle \mathrm{e}_d \wedge \mathrm{e}_e \wedge \mathrm{e}_f,\, \langle \omega^c,\, \mathrm{d}^2\, \mathrm{e}_a\, \omega^b \rangle\rangle \\ &= \tfrac{1}{3!}\, \langle \omega^c,\, \langle \mathrm{e}_d \wedge \mathrm{e}_e \wedge \mathrm{e}_f,\, \mathrm{d}^2\, \mathrm{e}_a\, \omega^b \rangle\rangle \\ &= \tfrac{1}{3!}\, \langle \mathrm{e}_d \wedge \mathrm{e}_e \wedge \mathrm{e}_f \otimes \omega^c,\, \mathrm{d}^2\, \mathrm{e}_a\, \omega^b \rangle \\ &= \tfrac{1}{3!}\, \langle \mathrm{e}_d \wedge \mathrm{e}_e \wedge \mathrm{e}_f\, \omega^c,\, \mathrm{d}^2\, \mathrm{e}_a\, \omega^b \rangle, \end{aligned} \tag{42}$$

where

$$\begin{aligned} R^{(3)b}{}_{defa}{}^c &= \delta^b_{[d}\, R_{ef]a}{}^c \\ &= \tfrac{1}{3}\, (\delta^b_d\, R_{efa}{}^c + \delta^b_e\, R_{fda}{}^c + \delta^b_f\, R_{dea}{}^c), \end{aligned} \tag{43}$$

whence $R^{(3)f}{}_{abcd}{}^e$ is given by

$$\begin{aligned} R^{(3)f}{}_{abcd}{}^e &= \delta^f_{[a}\, R_{bc]d}{}^e \\ &= \tfrac{1}{3}\, (\delta^f_a\, R_{bcd}{}^e + \delta^f_b\, R_{cad}{}^e + \delta^f_c\, R_{abd}{}^e) \\ &= \tfrac{1}{3!}\, \delta^{ghi}_{abc}\, \delta^f_g\, R_{hid}{}^e \\ &= \tfrac{1}{3}\, \langle \mathrm{e}_a \wedge \mathrm{e}_b \wedge \mathrm{e}_c,\, \Omega_d{}^e \wedge \omega^f \rangle \\ &= \tfrac{1}{3}\, \langle \mathrm{e}_a \wedge \mathrm{e}_b \wedge \mathrm{e}_c,\, \langle \omega^e,\, \mathrm{d}^2\, \mathrm{e}_d\, \omega^f \rangle\rangle \\ &= \tfrac{1}{3}\, \langle \omega^e,\, \langle \mathrm{e}_a \wedge \mathrm{e}_b \wedge \mathrm{e}_c,\, \mathrm{d}^2\, \mathrm{e}_d\, \omega^f \rangle\rangle \\ &= \tfrac{1}{3}\, \langle \mathrm{e}_a \wedge \mathrm{e}_b \wedge \mathrm{e}_c \otimes \omega^e,\, \mathrm{d}^2\, \mathrm{e}_d\, \omega^f \rangle \\ &= \tfrac{1}{3}\, \langle \mathrm{e}_a \wedge \mathrm{e}_b \wedge \mathrm{e}_c\, \omega^e,\, \mathrm{d}^2\, \mathrm{e}_d\, \omega^f \rangle, \end{aligned} \tag{44}$$

where $\delta^{ghi}_{abc}$ is the $6^{th}$-order generalized Kronecker delta.

The $4^{th}$-order contraction $R^{(3)f}{}_{abcf}{}^e$ of $R^{(3)f}{}_{abcd}{}^e$ is given by

$$\begin{aligned} R^{(3)f}{}_{abcf}{}^e &= \delta^d_f\, R^{(3)f}{}_{abcd}{}^e \\ &= \delta^d_f\, \delta^f_{[a}\, R_{bc]d}{}^e \\ &= \delta^f_{[a}\, R_{bc]f}{}^e \\ &= \tfrac{1}{3}\, (\delta^f_a\, R_{bcf}{}^e + \delta^f_b\, R_{caf}{}^e + \delta^f_c\, R_{abf}{}^e) \\ &= \tfrac{1}{3}\, (R_{bca}{}^e + R_{cab}{}^e + R_{abc}{}^e) \\ &= R_{[abc]}{}^e \\ &= 2\, (\nabla_{[a}\, S_{bc]}{}^e - 2\, S_{[ab}{}^f\, S_{c]f}{}^e). \end{aligned} \tag{45}$$

The $4^{th}$-order contraction $R^{(3)f}{}_{abc}{}^{ae}$ of $R^{(3)f}{}_{abcd}{}^e$ is given by

$$\begin{aligned} R^{(3)f}{}_{abc}{}^{ae} &= g^{ad}\, R^{(3)f}{}_{abcd}{}^e \\ &= g^{ad}\, \delta^f_{[a}\, R_{bc]d}{}^e \\ &= \delta^f_{[a}\, R_{bc]}{}^{ae} \\ &= \tfrac{1}{3}\, (\delta^f_a\, R_{bc}{}^{ae} + \delta^f_b\, R_{ca}{}^{ae} + \delta^f_c\, R_{ab}{}^{ae}) \\ &= g^{ad}\, g^{eg}\, \delta^f_{[a}\, R_{bc]dg} \\ &= g^{ad}\, g^{eg}\, (-\, \delta^f_{[a}\, R_{bc]gd} + 2\, \delta^f_{[a}\, R_{bc](dg)}) \\ &= -\, g^{ad}\, g^{eg}\, \delta^f_{[a}\, R_{bc]gd} + \\ &\quad + 2\, g^{ad}\, g^{eg}\, \delta^f_{[a}\, (\nabla_b\, Q_{c]dg} + S_{bc]}{}^h\, Q_{hdg}) \\ &= -\, g^{eg}\, \delta^f_{[a}\, R_{bc]g}{}^a + 2\, \delta^f_{[a}\, (\nabla_b\, Q_{c]}{}^{ae} + S_{bc]}{}^d\, Q_d{}^{ae}) \\ &= -\, \tfrac{1}{3}\, g^{eg}\, (\delta^f_a\, R_{bcg}{}^a + \delta^f_b\, R_{cag}{}^a + \delta^f_c\, R_{abg}{}^a) + \\ &\quad + 2\, \delta^f_{[a}\, (\nabla_b\, Q_{c]}{}^{ae} + S_{bc]}{}^d\, Q_d{}^{ae}) \\ &= -\, \tfrac{1}{3}\, g^{eg}\, (R_{bcg}{}^f - \delta^f_b\, R_{acg}{}^a + \delta^f_c\, R_{abg}{}^a) + \\ &\quad + 2\, \delta^f_{[a}\, (\nabla_b\, Q_{c]}{}^{ae} + S_{bc]}{}^d\, Q_d{}^{ae}) \\ &= -\, \tfrac{1}{3}\, g^{eg}\, (R_{bcg}{}^f - \delta^f_b\, R_{cg} + \delta^f_c\, R_{bg}) + \\ &\quad + 2\, \delta^f_{[a}\, (\nabla_b\, Q_{c]}{}^{ae} + S_{bc]}{}^d\, Q_d{}^{ae}) \\ &= -\, \tfrac{1}{3}\, g^{eg}\, (R_{bcg}{}^f - 2\, \delta^f_{[b}\, R_{c]g}) + \\ &\quad + 2\, \delta^f_{[a}\, (\nabla_b\, Q_{c]}{}^{ae} + S_{bc]}{}^d\, Q_d{}^{ae}) \\ &= -\, \tfrac{1}{3}\, (R_{bc}{}^{ef} - 2\, \delta^f_{[b}\, R_{c]}{}^e) + \\ &\quad + 2\, \delta^f_{[a}\, (\nabla_b\, Q_{c]}{}^{ae} + S_{bc]}{}^d\, Q_d{}^{ae}) \\ &= \tfrac{1}{3}\, [(-\, R_{bc}{}^{ef} + 2\, \delta^f_{[b}\, R_{c]}{}^e) + 2\, (\nabla_{[b}\, Q_{c]}{}^{fe} + S_{bc}{}^d\, Q_d{}^{fe}) + \\ &\quad + 2\, \delta^f_b\, (\nabla_{[c}\, Q_{a]}{}^{ae} + S_{ca}{}^d\, Q_d{}^{ae}) + \\ &\quad + 2\, \delta^f_c\, (\nabla_{[a}\, Q_{b]}{}^{ae} + S_{ab}{}^d\, Q_d{}^{ae})] \\ &= \tfrac{1}{3}\, [(R_{bc}{}^{fe} + 2\, \delta^f_{[b}\, R_{c]}{}^e) + \\ &\quad + 2\, \delta^f_b\, (\nabla_{[c}\, Q_{a]}{}^{ae} + S_{ca}{}^d\, Q_d{}^{ae}) + \\ &\quad + 2\, \delta^f_c\, (\nabla_{[a}\, Q_{b]}{}^{ae} + S_{ab}{}^d\, Q_d{}^{ae})], \end{aligned} \tag{46}$$

the $2^{nd}$-order contraction $R^{(3)f}{}_{abc}{}^{ab}$ of which is given by

$$R^{(3)f}{}_{abc}{}^{ab} = \delta^b_e\, R^{(3)f}{}_{abc}{}^{ae} \quad (47)$$
$$= \delta^b_e\, \delta^f_{[a}\, R_{bc]}{}^{ae}$$
$$= \delta^f_{[a}\, R_{bc]}{}^{ab}$$
$$= \tfrac{1}{3}\,(2\, G_c{}^f + \nabla_{[c}\, Q_{a]}{}^{fa} + S_{ca}{}^b\, Q_b{}^{fa})$$
$$= \tfrac{2}{3}\, G_c{}^f - \tfrac{1}{3}\,(\nabla_{[a}\, Q_{c]}{}^{af} + S_{ac}{}^b\, Q_b{}^{af}),$$

where

$$G_a{}^b = R_a{}^b - \tfrac{1}{2}\,\delta^b_a\, R \quad (48)$$

is the usual Einstein curvature tensor and $R_a{}^b$ and $R$ are the Ricci curvature tensor and Riemann curvature scalar of $M$ as given by

$$R_a{}^b = R_{ca}{}^{bc} \quad (49)$$

and

$$R = R_a{}^a = R_{ba}{}^{ab}, \quad (50)$$

respectively.

The 4$^{th}$-order contraction $R^{(3)f}{}_{abcd}{}^a$ of $R^{(3)f}{}_{abcd}{}^e$ is given by

$$R^{(3)f}{}_{abcd}{}^a = \delta^a_e\, R^{(3)f}{}_{abcd}{}^e \quad (51)$$
$$= \delta^a_e\, \delta^f_{[a}\, R_{bc]d}{}^e$$
$$= \delta^f_{[a}\, R_{bc]d}{}^a$$
$$= \tfrac{1}{3}\,(\delta^f_a\, R_{bcd}{}^a + \delta^f_b\, R_{cad}{}^a + \delta^f_c\, R_{abd}{}^a)$$
$$= \tfrac{1}{3}\,(R_{bcd}{}^f - \delta^f_b\, R_{acd}{}^a + \delta^f_c\, R_{abd}{}^a)$$
$$= \tfrac{1}{3}\,(R_{bcd}{}^f - \delta^f_b\, R_{cd} + \delta^f_c\, R_{bd})$$
$$= \tfrac{1}{3}\,(R_{bcd}{}^f - 2\,\delta^f_{[b}\, R_{c]d}),$$

the 2$^{nd}$-order contraction $R^{(3)f}{}_{abc}{}^{ba}$ of which is given by

$$R^{(3)f}{}_{abc}{}^{ba} = g^{bd}\, R^{(3)f}{}_{abcd}{}^a \quad (52)$$
$$= g^{bd}\, \delta^f_{[a}\, R_{bc]d}{}^a$$
$$= \delta^f_{[a}\, R_{bc]}{}^{ba}$$
$$= \tfrac{1}{3}\,(\nabla_{[b}\, Q_{c]}{}^{bf} + S_{bc}{}^e\, Q_e{}^{bf} + \delta^f_c\, R - 2\, R_c{}^f)$$
$$= \tfrac{1}{3}\,(\nabla_{[a}\, Q_{c]}{}^{af} + S_{ac}{}^b\, Q_b{}^{af} - 2\, G_c{}^f)$$
$$= -\tfrac{2}{3}\, G_c{}^f + \tfrac{1}{3}\,(\nabla_{[a}\, Q_{c]}{}^{af} + S_{ac}{}^b\, Q_b{}^{af}).$$

Thus,

$$G^{(3)f}{}_{abcf}{}^e = \tfrac{1}{2}\, R^{(3)f}{}_{abcf}{}^e \quad (53)$$
$$= \tfrac{1}{2}\, R_{[abc]}{}^e$$
$$= \nabla_{[a}\, S_{bc]}{}^e - 2\, S_{[ab}{}^f\, S_{c]f}{}^e,$$

$$G^{(3)f}{}_{abc}{}^{ab} = \tfrac{1}{2}\, R^{(3)f}{}_{abc}{}^{ab} \quad (54)$$
$$= \tfrac{1}{3}\, G_c{}^f - \tfrac{1}{6}\,(\nabla_{[a}\, Q_{c]}{}^{af} + S_{ac}{}^b\, Q_b{}^{af}),$$

and

$$G^{(3)f}{}_{abc}{}^{ba} = \tfrac{1}{2}\, R^{(3)f}{}_{abc}{}^{ba} \quad (55)$$
$$= -\tfrac{1}{3}\, G_c{}^f + \tfrac{1}{6}\,(\nabla_{[a}\, Q_{c]}{}^{af} + S_{ac}{}^b\, Q_b{}^{af}).$$

Equation (53) can be written

$$G^{(3)e}{}_{abce}{}^d = \tfrac{1}{2}\, R^{(3)e}{}_{abce}{}^d \quad (56)$$
$$= \tfrac{1}{2}\, R_{[abc]}{}^d$$
$$= \nabla_{[a}\, S_{bc]}{}^d - 2\, S_{[ab}{}^e\, S_{c]e}{}^d$$

and, using the result that

$$g_{fe}\,(\nabla_{[a}\, Q_{c]}{}^{af} + S_{ac}{}^b\, Q_b{}^{af}) = g_{fe}\, \nabla_{[a}\, Q_{c]}{}^{af} + g_{fe}\, S_{ac}{}^b\, Q_b{}^{af} \quad (57)$$
$$= \nabla_{[a}\, g_{|fe|}\, Q_{c]}{}^{af} - (\nabla_{[a}\, g_{|fe|})\, Q_{c]}{}^{af} +$$
$$+ S_{ac}{}^b\, Q_b{}^a{}_e$$
$$= \nabla_{[a}\, Q_{c]}{}^a{}_e + Q_{[a|fe|}\, Q_{c]}{}^{af} + S_{ac}{}^b\, Q_b{}^a{}_e,$$

Eqs. (54) and (55) can be written as

$$G^{(3)}{}_{bcda}{}^{cd} = \tfrac{1}{2}\, R^{(3)}{}_{bcda}{}^{cd} \quad (58)$$
$$= \tfrac{1}{3}\, G_{ab} - \tfrac{1}{6}\,(\nabla_{[c}\, Q_{a]}{}^c{}_b + S_{ca}{}^d\, Q_d{}^c{}_b + Q_{[c|bd|}\, Q_{a]}{}^{cd})$$
$$= \tfrac{1}{3}\, G_{ab} + \tfrac{1}{6}\,(\nabla_{[a}\, Q_{c]}{}^c{}_b + S_{ac}{}^d\, Q_d{}^c{}_b + Q_{[a|bd|}\, Q_{c]}{}^{cd})$$

and

$$G^{(3)}{}_{bcda}{}^{dc} = \tfrac{1}{2}\, R^{(3)}{}_{bcda}{}^{dc} \quad (59)$$
$$= -\tfrac{1}{3}\, G_{ab} + \tfrac{1}{6}\,(\nabla_{[c}\, Q_{a]}{}^c{}_b + S_{ca}{}^d\, Q_d{}^c{}_b + Q_{[c|bd|}\, Q_{a]}{}^{cd})$$
$$= -\tfrac{1}{3}\, G_{ab} - \tfrac{1}{6}\,(\nabla_{[a}\, Q_{c]}{}^c{}_b + S_{ac}{}^d\, Q_d{}^c{}_b + Q_{[a|bd|}\, Q_{c]}{}^{cd}),$$

respectively, whence

$$R_{[abc]}{}^d = 2\,(\nabla_{[a}\, S_{bc]}{}^d - 2\, S_{[ab}{}^e\, S_{c]e}{}^d) \quad (60)$$
$$= R^{(3)e}{}_{abce}{}^d$$
$$= 2\, G^{(3)e}{}_{abce}{}^d,$$

$$G_{ab} = \tfrac{3}{2}\, R^{(3)}{}_{bcda}{}^{cd} - \tfrac{1}{2}\,(\nabla_{[a}\, Q_{c]}{}^c{}_b + S_{ac}{}^d\, Q_d{}^c{}_b + Q_{[a|bd|}\, Q_{c]}{}^{cd}) \quad (61)$$
$$= 3\, G^{(3)}{}_{bcda}{}^{cd} - \tfrac{1}{2}\,(\nabla_{[a}\, Q_{c]}{}^c{}_b + S_{ac}{}^d\, Q_d{}^c{}_b + Q_{[a|bd|}\, Q_{c]}{}^{cd}),$$

and

$$G_{ab} = -\tfrac{3}{2}\, R^{(3)}{}_{bcda}{}^{dc} - \tfrac{1}{2}\,(\nabla_{[a}\, Q_{c]}{}^c{}_b + S_{ac}{}^d\, Q_d{}^c{}_b + Q_{[a|bd|}\, Q_{c]}{}^{cd}) \quad (62)$$
$$= -3\, G^{(3)}{}_{bcda}{}^{dc} - \tfrac{1}{2}\,(\nabla_{[a}\, Q_{c]}{}^c{}_b + S_{ac}{}^d\, Q_d{}^c{}_b + Q_{[a|bd|}\, Q_{c]}{}^{cd}).$$

Thus, the equations

$$G^{(3)f}{}_{abcd}{}^e = \tfrac{1}{2}\, R^{(3)f}{}_{abcd}{}^e \quad (63)$$
$$= \tfrac{1}{2}\, \delta^f_{[a}\, R_{bc]d}{}^e$$
$$= \tfrac{1}{3}\, k\, T^{(3)f}{}_{abcd}{}^e,$$

where $k$ is a constant, which may be taken to be given by

$$k = \frac{8\pi G}{c^2}, \quad (64)$$

where $G$ is Newton's gravitational constant and $c$ is the speed of light in vacuum, and where

$$T^{(3)b}{}_{cda}{}^{cd} = -\, T^{(3)b}{}_{cda}{}^{dc} \quad (65)$$
$$= T_a{}^b$$

or equivalently

$$T^{(3)}{}_{bcda}{}^{cd} = -\, T^{(3)}{}_{bcda}{}^{dc} \quad (66)$$
$$= g_{be}\, T^{(3)e}{}_{cda}{}^{cd}$$
$$= -\, g_{be}\, T^{(3)e}{}_{cda}{}^{dc}$$
$$= T_{ab}$$

and

$$T^{(3)e}{}_{abce}{}^d = 0, \quad (67)$$

where $T_{ab}$ is the 5-dimensional matter tensor comprising the 4-dimensional matter tensor and the 4-dimensional electric current vector (in addition to a hypothetical but otherwise unspecified source term), can be interpreted—aside from the additional, "correction" terms involving $S_{ab}{}^c$ and $Q_a{}^{bc}$—as the usual Einstein-Maxwell field equations as given in 5-dimensional general relativity in particular by

$$G_{ab} = R_{ab} - \tfrac{1}{2}\, g_{ab}\, R \quad (68)$$
$$= k\, T_{ab},$$

which correspond to the usual non-source-free (or "source" or "inhomogeneous") Einstein field equations (moreover, automatically including the usual electromagnetic contributions to the Ricci curvature tensor and Riemann curvature scalar) and the usual non-source-free (or "source," "divergence," or "inhomogeneous") Maxwell field equations (in addition to a possible 5-dimensional non-source-free generalization of Schrödinger's equation[38] or of the usual Klein-Gordon equation[39-41]), together with

$$\begin{aligned} R_{[abc]}{}^{d} &= \tfrac{1}{3}\,(R_{abc}{}^{d} + R_{bca}{}^{d} + R_{cab}{}^{d}) \qquad (69) \\ &= 2\,(\nabla_{[a}\, S_{bc]}{}^{d} - 2\, S_{[ab}{}^{e}\, S_{c]e}{}^{d}) \\ &= 2\,(\nabla_{[a}\, S_{bc]}{}^{d} + 2\, S_{[ab}{}^{e}\, S_{|e|c]}{}^{d}) \\ &= -\, g^{de}\, R_{[ab|e|c]} + 2\, g^{de}\,(\nabla_{[a}\, Q_{bc]e} + S_{[ab}{}^{f}\, Q_{|f|c]e}) \\ &= -\, R_{[ab}{}^{d}{}_{c]} + 2\, \nabla_{[a}\, g^{de}\, Q_{bc]e} - 2\,(\nabla_{[a}\, g^{de})\, Q_{bc]e} + \\ &\quad + 2\, S_{[ab}{}^{f}\, g^{de}\, Q_{|f|c]e} \\ &= -\, R_{[ab}{}^{d}{}_{c]} + 2\, \nabla_{[a}\, Q_{bc]}{}^{d} - 2\, Q_{[a}\, g^{de}\, Q_{bc]e} + 2\, S_{[ab}{}^{e}\, Q_{|e|c]}{}^{d} \\ &= 0, \end{aligned}$$

which correspond to the usual source-free (or "rotational" or "homogeneous") Maxwell field equations[42-49] and which also—along with Eq. (67) for $T^{(3)e}{}_{abce}{}^{d}$—can be modified to account for certain as yet hypothetical magnetic source terms as needs be.

---

38 Gonseth, F., and G. Juvet, "Sur la relativité à cinq dimensions et sur une interprétation de l'équation de Schrödinger," *Helvetica Physica Acta*, **1** (1928) 421; see also "Sur les équations de l'électromagnétisme," *Comptes rendus hebdomadaires des séances de l'Académie des Sciences (Paris)*, **185** (1927) 341, and errata, *ibid.*, 483; "Sur la métrique de l'espace à 5 dimensions de l'électromagnétisme et de la gravitation," *ibid.*, 412; "Sur l'équation de M. Schrödinger," *ibid.*, 448; "Les équations de l'électromagnétisme et l'équation de M. Schrödinger dans l'univers à cinq dimensions," *ibid.*, 535, and errata, *ibid.*, 732.

39 Liu, H., and P. S. Wesson, "Kaluza-Klein-Gordon Field Theory," *Int. Journ. Mod. Phys. D*, **7** (1998) 737.

40 Sajko, W. N., P. S. Wesson, and H. Liu, "Waves and particles in Kaluza-Klein theory," *J. Math. Phys.*, **40** (1999) 2364.

41 Wesson, P. S., *Space, Time, Matter: Modern Kaluza-Klein Theory*, World Scientific Publishing Co. Pte. Ltd., Singapore (1999), pp. 196-197.

42 Kaluza, T., "Zum Unitätsproblem der Physik," *Sitzungsberichte der preussischen Akademie der Wissenschaften*, **54** (1921) 966.

43 Mandel, H., "Zur Herleitung der Feldgleichungen in der allgemeinen Relativitätstheorie. (Zweite Mitteilung.)," *Zeitschrift für Physik*, **45** (1927) 285.

44 Rumer, G., "Über eine Erweiterung der allgemeinen Relativitätstheorie," *Nachrichten von der Gesellschaft der Wissenschaften zu Göttingen, Mathematisch-physikalische Klasse*, **2** (1929) 92; "Form und Substanz," *Zeitschrift für Physik*, **58** (1929) 273.

45 Miron, R., and M. Radivoiovici-Tatoiu, "Extended Lagrangian Theory of Electromagnetism," *Rep. Math. Phys.*, **27** (1989) 193.

46 Kawaguchi, T., and R. Miron, "A Lagrangian Model for Gravitation and Electromagnetism," *Tensor (N. S.)*, **48** (1989) 153.

47 Miron, R., R. Rosca, M. Anastasiei, and K. Buchner, "New Aspects of Lagrangian Relativity," *Found. Phys. Lett.*, **5** (1992) 141.

48 Schwarz, A. J., and N. A. Doughty, "Kaluza-Klein unification and the Fierz-Pauli weak-field limit," *Am. J. Phys.*, **60** (1992) 150.

49 Ringermacher, H. I., and B. N. Cassenti, "An Experiment to Verify a Relativistic Field Theory Based on an Electrodynamic Connection," in *Proceedings of the 34th AIAA/ASME/SAE/ASEE Joint Propulsion Conference & Exhibit, July 13-15, 1998/Cleveland, OH*, American Institute of Aeronautics and Astronautics, Washington, DC (1998), paper no. AIAA-98-3136.